\documentstyle[12pt]{article}
\begin{document}

\baselineskip=22pt plus 0.2pt minus 0.2pt
\lineskip=22pt plus 0.2pt minus 0.2pt

\begin{center}
\large

Quark Delocalization, Color Screening, and Nuclear \\
Intermediate Range Attraction\\

\vspace*{0.25in}

Fan Wang,$^{1)}$ Guang-han Wu,$^{2)}$,\\
Li-jian Teng,$^{2)}$ and T.\ Goldman,$^{3)}$

\end{center}

\vspace*{0.20in}
$^{1)}$ Center for Theoretical Physics, Nanjing University, Nanjing, China,
210008 and CCAST (World Laboratory)

$^{2)}$ Institute of Nuclear Science and Technology, Sichuan University,
Cheng-du, China, 610064

$^{3)}$ Theoretical Division, Los Alamos National Laboratory, NM 87545, USA

\vspace*{0.20in}

\begin{center}
\large

\vspace*{1.25in}

Abstract

\end{center}

\begin{center}

\noindent We consider the effect of including quark delocalization
and color screening, in the nonrelativistic quark cluster model, on
baryon-baryon potentials and phase shifts. We find that the inclusion
of these additional effects allows a good qualitative description of
both.

\end{center}

\newpage

\noindent {\bf I. Introduction}

Up to now, all quark model calculations (including potential, bag, and
soliton models) of the nuclear force have found difficulty in obtaining
the observed intermediate range attraction.$^{[1]}$  It is not even
technically easy for meson exchange models$^{[2]}$ to do so.
Therefore, one should consider the possibility that all of these models
have omitted important physical effects.

In molecular physics, experience shows that electron delocalization is
an important effect contributing to the formation of chemical bonds.
Is there a similar effect for the nuclear bond due to
quark delocalization?  (This effect has also been called quark
percolation in view of the fact that quarks are confined in color
singlet hadrons.)

As in molecular physics, there are also other forces at work.  There,
multiphoton exchange produces (van der Waals') interactions between
neutral atoms. But there is no evidence for (color van der Waals')
multigluon exchange forces between color singlet hadrons. So, although
inside a hadron, quarks experience a confining interaction, between two
colorless hadrons, the confinement interaction must certainly be
modified. Lattice gauge calculations have indeed shown qualitatively
that there is a color screening effect due to $q\overline{q}$
excitation$^{[3]}$. How important color screening is to the nuclear
interaction remains an open question.

We have made a (nonrelativistic) QCD model calculation to study whether
a nuclear intermediate range attraction arises when we take into
account the possibility of both quark delocalization and color
screening.  The result we find is that quark delocalization and color
screening actually do seem to give rise to an effect very similar to
the nuclear intermediate range attraction.  It seems that quark
delocalization and color screening may play an effective role similar
to meson exchange.


\noindent{\bf II. The Quark Delocalization and Color Screening Model}

We take a nonrelativistic quark cluster model to generate our QCD model
for the baryon-baryon interaction. The new ingredients are quark
delocalization and color screening. For individual baryons, our model
is the same as the usual one$^{[4]}$. That is we have the Hamiltonian
$$
H = \sum_{i} (m_{i} + \frac{P^{2}_{i}}{2m_{i}}) -T_{c} + \sum_{i<j} V_{ij}
\eqno(1)
$$
$$
V_{ij} = V^{C}_{ij} + V^{G}_{ij} \eqno(2a)
$$
$$
V^{C}_{ij} = -\vec{\lambda}_{i} \cdot \vec{\lambda}_{j}  ar^{2} \eqno(2b)
$$
$$
V^{G}_{ij} = \alpha_{S} \frac{\vec{\lambda}_{i} \cdot \vec{\lambda}_{j}}{4}
\left[ \frac{1}{r} - \frac{\pi}{2} \delta(\vec{r})(\frac{1}{m_{i}} +
\frac{1}{m_{j}} + \frac{4}{3m_{i}m_{j}} \vec{\sigma}_{i} \cdot
\vec{\sigma}_{j})\right] \eqno(2c)
$$
\noindent where $\vec{r} = \vec{r}_{i} - \vec{r}_{j}$ and $T_{c}
\equiv$ center of mass kinetic energy. Here we keep only the effective
one gluon exchange form of the color Coulomb and hyperfine terms and
also neglect a possible constant part of the confining interaction. We
do so because we are interested in the qualitative features of this
model and wish to keep the number of parameters to a minimum. It should
be noted, however, that our form for $V_{ij}$ is consistent with all
charmonium and Upsilon bound state spectra.$^{[5]}$

The single quark orbital wave function is chosen to be a Gaussian
function
$$
\psi (\vec{r}_{i}) = (\frac{1}{\pi b^{2}})^{3/4} \exp\{-
\frac{1}{2b^{2}}(\vec{r}_{i} - \vec{R})^{2} \} \eqno(3)
$$
\noindent where $\vec{R}$ is the reference center (mean location of the
baryon).  The parameters are determined as follows. We choose the
constituent quark mass to be exactly 1/3 of the nucleon mass, i.e. $m =
313 MeV$, so that the binding energy of the nucleon is exactly zero. We
also require the model to produce the correct $N-\Delta$ mass
difference and that the nucleon size satisfies the stability condition,
i.e.  ${\delta M_{N}}/{\delta b} = 0$. The fitted model parameters
are:
$$
m = 313 MeV, \;\; b= 0.603 fm, \;\; \alpha_{S} = 1.54, \;\; a =
25.13MeV/fm^{2} \eqno(4)
$$
\noindent These are similar to the parameters chosen in Ref.\ [4].

For the two baryon system, we take a two ``center'' cluster model
approximation, i.e., we have the left (right) single quark orbital
wavefunction (w.\ f.\ )
$$
\phi_{L}(\vec{r}_{i}) = (\frac{1}{\pi b^{2}})^{3/4} \exp\{-\frac{1}{2b^{2}}
(\vec{r}_{i} + \vec{R}/2)^{2} \} \eqno(5a)
$$
$$
\phi_{R}(\vec{r}_{i}) = (\frac{1}{\pi b^{2}})^{3/4} \exp
\{- \frac{1}{2b^{2}}(\vec{r}_{j} - \vec{R}/2)^{2} \} \eqno(5b)
$$
\noindent Now $\vec{R}$ is the distance between the two clusters.

The variational trial w.f.\ of the two baryon system is an
antisymmetric six quark product state
$$
\Psi (B_{1}B_{2}) = {\cal A} \{ [
\psi_{L}(1)\psi_{L}(2)\psi_{L}(3)]_{B_{1}}[\psi_{R}(4)
\psi_{R}(5)\psi(6))_{B_{2}}] \} _{ST}  \eqno(6)
$$
\noindent $\cal A$ is the antisymmetrization operator, $[\;\;]_{B}$
means the spin, isospin, and color of the three quarks are coupled to
the quantum numbers of a baryon, $\{\;\;\}_{ST}$ means the spin,
isospin and color of the two baryons are coupled to the particular
color singlet channel of spin $S$ and isospin $T$.

One of our new ingredients, quark delocalization, is put in through the
trial single quark w.\ f.\ $\psi$
$$
\psi_{L} = \phi_{L} + \epsilon \phi_{R}, \;\; \psi_{R} = \phi_{R} + \epsilon
\phi_{L} \eqno(7)
$$
\noindent where $\epsilon = \epsilon(\vec{R})$ (see below) is a
variational parameter determined by a variational calculation, after
the manner suggested in Ref.\ [6] in a relativistic quark picture of
nuclear structure. We explicitly examine here only the $\epsilon > 0$
region.  Numerical results for negative $\epsilon$ show increased
energy for the state. This confirms expectations based on viewing this
case as including $\;p-$wave contributions of necessarily higher energy,
or from the point of view of two-state mixing (producing symmetric and
antisymmetric states in the $\epsilon = \pm 1$ limit).

The other new ingredient, color screening, is put in through the
modification of the confinement interaction between two color singlet
baryons
$$
V^{SC}_{ij} = - \vec{\lambda}_{i} \cdot \vec{\lambda}_{j} a r^{2} e^{-\mu
r^{2}} \eqno(8)
$$
\noindent When we calculate the six quark confinement interaction
matrix elements, we separate them into internal and inter-cluster
parts: for the latter we use the screening confinement interaction
$V^{SC}_{ij}$.  Note that $\mu$ is a parameter which will be fixed by
data. However, since we have alluded to screening (found in lattice
calculations) due to $q\overline{q}$ production, and the scale for that
is set by the pion mass, we expect $\mu \sim m^{2}_{\pi}$ to result.

We do two kinds of calculations. First we calculate the diagonal matrix
element
$$
\left.
\frac{<\Psi_{B_{1}B_{2}}(\vec{R})|H|\Psi_{B_{1}B_{2}}(\vec{R})>}{<\Psi_{B_{1
}B_{2
}}(\vec{R})|\Psi_{B_{1}B_{2}}(\vec{R})>} -
\frac{<\Psi_{B_{1}B_{2}}|H|\Psi_{B_{1}B_{2}}>}{<\Psi_{B_{1}B_{2}}|\Psi_{B_{1
}
B_{2}}>} \right|_{\vec{R} = \infty} \eqno(9)
$$
\noindent to get the adiabatic approximation to the baryon-baryon
interaction.  For fixed $\mu$ at each separation $\vec{R}$, we vary the
parameter $\epsilon(\vec{R})$ to get the minimum. (We then vary $\mu$,
repeating the calculation until we get a best fit for the $NN \;
^{1}S_{0}$ phase shift, i.\ e.\ minimum $\chi^2$ per degree of freedom;
see below). We take this as the approximation of our model
baryon-baryon interaction.

Second, we do a dynamical calculation to get the phase shift of
baryon-baryon scattering.  When we do this calculation, we have to
calculate the off-diagonal kernal in addition to the diagonal one:
$$
<\Psi_{B_{1}B_{2}}(\vec{R})|H| \Psi_{B_{1}B_{2}}(\vec{R}')> \eqno(10)
$$
\noindent For $\Psi(\vec{R})$ we use the parameter $\epsilon = \epsilon
(\vec{R})$ and for $\Psi(\vec{R}')$ we use $\epsilon = \epsilon
(\vec{R}')$. Finally we use the Canto-Brink$^{[7]}$ (CB) variational
method to calculate the phase shift and vary $\mu$ to get a best fit to
$NN \; ^{1}S_{0}$ phase shift. The CB method is equivalent to the more
popular resonating group method (RGM).$^{[8]}$

\noindent {\bf III. Results and Discussion}

We have calculated the effective baryon-baryon interaction and the
phase shifts for $NN \;\; (S,T) = (1,0), (0,1)$ and $\Delta \Delta \;\;
(S,T) = (3,0)$ channels. The results are shown in Figs.\ 1-2.  The
experimental phase shifts are taken from Table IV of Ref.\ [9]. The
optimum color screening constant is $\mu = 0.46 fm^{-2}$, consistent
with our expectations. The results are not very sensitive to $\mu$
within the range 0.4 to 0.5 fm$^{-2}$, and an intermediate attraction
persists over an even wider range of values.  (It would be interesting
to compare this with meson exchange results over a range of pion mass
values.) The fitted quark delocalization parameters $\epsilon$ are
given in Table 1.

It is well known that the pure quark cluster with gluon exchange model
can only give rise to a $N-N$ repulsive core.  Figs.\ 1 and 2 however,
show that quark delocalization and color screening working together can
produce a good description of the $N-N$ interaction: it has both the
repulsive core {\it and} an intermediate range attraction.  Further,
the $NN \;\; ^{3}S_{1}$ channel attraction is stronger than $^{1}S_{0}$
channel.  Qualitatively the result fits the $^{1}S_{0}$ channel phase
shifts; it also gives a qualitative fit for the $^{1}D_{2}$ phase
shifts.  The fit to the $^{3}S_{1}$ phase shifts is not as good as
$^{1}S_{0}$, and $^{1}D_{2}$.  This is not unreasonable, because we
have not yet included the tensor coupling.

	This model also automatically gives a reasonable amount of
delocalization.  When the two nucleons are close together, the
delocalization is large; $\epsilon (R=0.375fm) \sim 1$, which means
that at short distances, the six quarks prefer to merge into a six
quark state instead of two nucleons.$^{[10,11]}$  However, when the two
nucleons separate to a distance comparable to the average distance
between nucleons in a nucleus, the delocalization is small:
$\epsilon(R = 1.5 fm)\sim 0.1$, which means the six quarks prefer to be
confined in two individual nucleons.$^{[6]}$

	It is worth mentioning here that the quark delocalization and
color screening effects must both be included; dropping either one
causes the $NN$ intermediate attraction to disappear.  However the $NN$
intermediate range attraction mainly comes from the kinetic energy
reduction due to quark delocalization.  It is only somewhat concealed
by the screened confining interaction.  If we drop the color screening,
however, the confinement interaction contribution related to the quark
delocalization will cancel the attraction coming from the kinetic
energy reduction related to the same quark delocalization.

	It seems likely that the effective attraction due to quark
kinetic energy reduction, which in turn comes from quark
delocalization, is quite general for all quark systems.  To show this,
we have also calculated the adiabatic interaction potential and phase
shifts of the $\Delta \Delta \;(S, T, \ell = 3,0,0)$ channel.  We find
that it is indeed an ``inevitable'' dibaryon state; our new
nonrelativistic quark cluster model also gives a strong intermediate
attraction, which supports the original LAMP calculation.$^{[12]}$
The reason for a stronger $\Delta \Delta$ attraction than that of $NN$
channels is due to a reduction in the cancellation of the attraction
due to quark delocalization by the screening confinement interaction in
the $\Delta \Delta \;\; (S, T, \ell = 3,0,0)$ channel.

	Note that there is no color van der Waals' force problem
for this model because we have taken color screening into account; any
such forces become gaussianly suppressed.  Whether this model also
gives a bound ``H''-state particle is an interesting question; a three
channel coupling calculation is in progress.  At the same time, we will
check if this model can give a reasonable deuteron bound state, which
would imply that the agreement with the phase shifts extends to very
low energy.

	Several points remain to be studied:
\begin{enumerate}
\item Due to the quark delocalization, the center of mass wave function
of this six quark system can not be separated in as clear cut a way as
in the usual cluster model.  To develop a method to separate the center
of mass wave function explicitly would be highly desirable.

\item We have given an intuitive argument for quark delocalization and
color screening. A better basis in QCD is needed.

\item This model seems to imitate meson exchange to some extent, but
the detailed relation between quark delocalization and meson exchange
is not yet clear.

\item The model qualitatively fits both hadron spectroscopy and the
$NN$ interaction, and quark delocalization seems to be a general
feature for all quark systems.  Therefore it would be interesting to
confront this model with possible dibaryon resonances.

\end{enumerate}

	We are happy to thank Jerry Stephenson, Kim Maltman, Karl
Holinde and Joe Carlson for valuable discussions.  We also thank LAMPF,
TRIUMF, and the Institute for Nuclear Theory (Seattle), where parts of
this work were conceived, for their hospitality.

	This work was supported in part by the National Science
Foundation of China and the US DOE.

\newpage

\noindent{\bf References}

\begin{enumerate}
\item Y.\ Fujiwara and K.\ T.\ Hecht, {\it Nucl.\ Phys.}
{\bf A444}, 541 (1985) and references therein.
\item R.\ Machleidt, K.\ Holinde and Ch.\ Elster, {\it Phys.\ Rep.}
{\bf 149}, 1 (1987).
\item B.\ Svetitsky in {\it ``Nuclear Chromodynamics''}, ed. by S.\ Brodsky
and E.\ Moniz (World Scientific, Singapore, 1986) pp.80-81.
\item N.\ Isgur and G.\ Karl, {\it Phys.\ Rev.\ D} {\bf 20}, 1191 (1979).
\item W.\ Kwong, J.\ L.\ Rosner and C.\ Quigg, {\it Ann.\ Rev.\ Nucl.\
Part.\ Sci.} {\bf 37}, 325 (1987).
\item T.\ Goldman, K.\ Maltman, G.\ J.\ Stephenson, Jr.\ and K.\ E.\ Schmidt,
{\it Nucl.\ Phys.} {\bf A481}, 621 (1988).
\item L.\ F.\ Canto and D.\ M.\ Brink, {\it Nucl.\ Phys.}
{\bf A279}, 85 (1977).
\item G.\ H.\ Wu and D.\ L.\ Yang, {\it Scientia Sinica, Suppl. A},
1112 (1982).
\item R.\ A.\ Arndt, {\it et al}., {\it Phys.\ Rev.\ D} {\bf 28}, 97 (1983).
\item G.\ E.\ Brown and M.\ Rho, {\it Phys.\ Lett.} {\bf 82B}, 177 (1979).
For a full discussion, see Y.\ He, F.\ Wang, and C.\ W.\ Wong, {\it Nucl.
Phys.} {\bf A448}, 652 (1986).
\item E.\ M.\ Henley, L.\ S.\ Kisslinger, and G.\ A.\ Miller, {\it Phys.\
Rev.\ C} {\bf 28}, 1277 (1983).
\item T.\ Goldman, K.\ Maltman, G.\ J.\ Stephenson, Jr.\ , K.\ E.\ Schmidt,
and F.\ Wang {\it Phys.\ Rev.\ C} {\bf 39}, 1889 (1989).
\end{enumerate}

\newpage
\vspace*{0.20in}

\begin{center}
Table 1.  Quark Delocalization Parameters $\epsilon(\vec{R})$\\
\begin{tabular}{lrrrrrrrrr}
R(fm) & 0.375 & 0.750 & 1.125 & 1.500 & 1.875 & 2.250 & 2.625 & 3.000\\
$NN \;\; ^{1}S_{0}$ & 0.999 & 0.510 & 0.166 & 0.123 & 0.077 & 0.038 & 0.015
&
0.005\\
$NN \;\; ^{3}S_{1}$ & 0.999 & 0.994 & 0.229 & 0.155 & 0.093 & 0.045 & 0.017
&
0.005\\
$\Delta \Delta \;\; ^{7}S_{3}$ & 0.999 & 0.999 & 0.999 & 0.999 & 0.999 &
0.999 &
0.208 & 0.055\\
\end{tabular}
\end{center}

\newpage
\vspace*{0.20in}

\noindent{\bf Figure Captions}

Figure 1. Baryon-baryon potentials as a function of separation in the
quark cluster model with quark delocalization and color screening;
a) two $NN$ channels, b) a $\Delta\Delta$ channel, c) detail of a) in
region of minimum.

Figure 2. Nucleon-nucleon phase shifts as a function of beam kinetic
energy in the quark cluster model with quark delocalization and color
screening, and the corresponding data from Table IV of Ref.\ [9]. The
parameters have been chosen to best fit the $^{1}S_{0}$ channel. Note
that tensor forces, which affect only the $^{3}S_{1}$ channel here,
have not been included.

\end{document}